\documentclass[letterpaper, 10 pt, conference]{IEEEtran}

\usepackage{amsmath, graphicx}
\usepackage{amssymb}

\title{The Redundancy of Slepian-Wolf Coding Revisited}
\author{\IEEEauthorblockN{Duo Xu}
\IEEEauthorblockA{Department of Electrical and Computer Engineering \\
University of Waterloo, ON, N2L 3G1, Canada \\
 Email: \{d35xu\}@uwaterloo.ca}}

\begin{document}
\maketitle

\begin{abstract}
[Draft] In this paper, the redundancy of Slepian Wolf coding is revisited. Applying the random binning and converse technique in \cite{yang}, the same results in \cite{he} are obtained with much simpler proofs. Moreover, our results reflect more details about the high-order terms of the coding rate. The redundancy is investigated for both fixed-rate and variable-rate cases. The normal approximation (or dispersion) can also be obtained with minor modification.
\end{abstract}

\section{Introduction}
This paper investigates the distributed source coding problem where two random variables $X$ and $Y$ are encoded separately but only $X$ is to be recovered. The scenario of the coding problem is depicted in Fig.1. The encoder observes and encodes only the source sequence $X^n$ generated by the memoryless source $X$, and then transmits the codeword to the decoder. The decoder observes both the transmitted codeword and the side information sequence $Y^n$ generated by a memoryless source $Y$, which is correlated with $X$. Our goal is to determine how many bits $R$ are required at least to decode $X$ with desired average error probability. 

This problem here is actually a special case of the distributed source coding problem proposed by Slepian and Wolf in \cite{SW}. It is now commonly referred to as the Slepian-Wolf (SW) coding problem (with one encoder).  Based on the Slepian Wolf coding theorem, under this scenario, $X$ can be recovered with arbitrarily small error probability $\epsilon$ when the rate $R$ is {\em asymptotically} close to $H(X|Y)$.

SW coding has great impact on the network information theory. So many papers made significant contribution to the practical SW code design \cite{xiong}, \cite{ramch}, \cite{rose}. In this case, the code blocklength $n$ and the error probability $\epsilon$ should be reasonably small. So, it is necessary to investigate the {\em redundancy} of the SW coding, which reflects how quickly the coding rate can approach the fundamental limitation as $n$ increases. We define the redundancy of SW coding $C_n$ as $r(C_n)-H(X|Y)$, where $r(C_n)$ is the average compression rate.

In \cite{he}, authors show that the redundancy of SW coding is $d_f\sqrt{-\log\epsilon_n/n}+o(\sqrt{-\log\epsilon_n/n})$ for the fixed rate case and $d_v\sqrt{-\log\epsilon_n/n}+o(\sqrt{-\log\epsilon_n/n})$ for the variable rate case. Recently, Yang etc. proposed the NEP theorem in \cite{nep}. In this paper, with the help of NEP theorem, the proof of the redundancy of SW coding can be greatly simplified. Especially, in the converse, with the converse technique \cite{yang}, the proof can be greatly simplified.

Recently, the author in \cite{japan1} and \cite{japan2} also proposed a simplified proof for the achievability of the redundancy, which was not able to show the converse.

This paper is organized as follows. In Section II, some notations and existing theorems which will be used in the proof are introduced. And in Section III, we derive the achievability for variable-rate and fixed-rated cases respectively. The converse is shown in Section IV. 

\begin{figure}[htb]
  \centerline{\includegraphics[width=5.5cm,height=3cm]{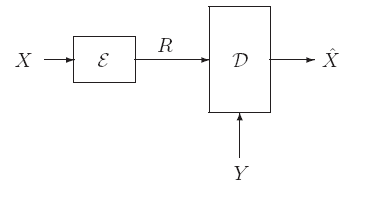}} 

  \caption{\setlength{\baselineskip}{0pt}SW coding scenario.}
  \label{fig:fig1}
\end{figure}

\section{Preliminary}
\subsection{Notation}
Let $(X,Y)$ be a pair of random variables with joint distribution $P_{XY}$ and finite conditional entropy $H(X|Y)$ measured on nats. We assume that both $X$ and $Y$ are discrete, and take values on the finite alphabets $\mathcal{X}$ and $\mathcal{Y}$. Let $p(x|y)$ be the conditional pmf of $X$ given $Y$, and $p(y)$ the pmf of $Y$. 

Let $\mathcal{P}(\mathcal{X}\times\mathcal{Y})$ denote the set of all distributions over $\mathcal{X}\times\mathcal{Y}$. And $\mathcal{P}^+(\mathcal{X}\times\mathcal{Y})$ denotes the subset of $\mathcal{P}(\mathcal{X}\times\mathcal{Y})$ with distributions with zero entries excluded. Let $\mathcal{T}_n(\mathcal{X})$ denote the set of all types on $\mathcal{X}^n$, and $\tau(x^n)$ be the type of $x^n$. Moreover, for any $t\in\mathcal{T}_n(\mathcal{X})$, let $T_{\mathcal{X}}^n(t)$ be the set of all $x^n$ with type $t$. Define
\begin{equation}
H(t)=\sum_{x\in\mathcal{X}}t(x)\ln\frac{1}{t(x)} \nonumber
\end{equation}
We denote $t\circ P_{Y|X}$ as the joint distribution of $X$ and $Y$ where $X$ obeys the distribution $t$. And $(t\circ P_{Y|X})_{\mathcal{Y}}$ represents the marginal of $t\circ P_{Y|X}$ over $\mathcal{Y}$.
\subsection{NEP With Respect To Conditional Entropy}
Define
\begin{eqnarray}
\sigma^2_H(X|Y)&\triangleq&\sum_{x\in\mathcal{X}}\sum_{y\in\mathcal{Y}} p(y)p(x|y)[-\ln p(x|y)]^2 \nonumber \\
\label{eq1}
&&-H^2(X|Y)
\end{eqnarray}
Further assume that $\sigma_H(X|Y)>0$. 
Define for any $\delta\ge0$
\begin{eqnarray}
\label{eq2}
r_{X|Y}(\delta)&\triangleq&\sup_{\lambda\ge0}\bigg[\lambda(H(X|Y)+\delta) \nonumber \\
&&-\ln\sum_{x\in\mathcal{X}}\sum_{y\in\mathcal{Y}}p(y)p^{-\lambda+1}(x|y)\bigg]
\end{eqnarray}
The right NEP with respect to $H(X|Y)$ in \cite{nep} is presented as below. \\
{\bfseries Theorem 1:} For any positive integer $n$,
\begin{equation}
\label{eq3}
\Pr\left\{-\frac{1}{n}\ln p(X^n|Y^n)>H(X|Y)+\delta\right\}\le e^{-nr_{X|Y}(\delta)}
\end{equation}
where $X^n=X_1X_2\ldots X_n$ and $Y^n=Y_1Y_2\ldots Y_n$. There exists a $\delta^*>0$ such that for any $\delta\in(0,\delta^*]$ and any positive integer $n$,
\begin{equation}
\label{eq4}
r_{X|Y}(\delta)=\frac{1}{2\sigma^2_H(X|Y)}\delta^2+O(\delta^3)
\end{equation}
and hence 
\begin{eqnarray}
&&\Pr\left\{-\frac{1}{n}\ln p(X^n|Y^n)>H(X|Y)+\delta\right\} \nonumber \\
\label{eq5}
&\le&e^{-n(\frac{\delta^2}{2\sigma^2_H(X|Y)}+O(\delta^3))}
\end{eqnarray}
\subsection{NEP With Respect to Relative Entropy}
Consider an IID source pair $(X,Y)=\{(X_i,Y_i)\}
_{i=1}^{\infty}$ with finite mutual information $I(X,Y)>0$. For any $t\in\mathcal{P}$, let
\begin{eqnarray}
\label{eq6}
q_t(y)&\triangleq&\sum_{x\in\mathcal{X}}t(x)p(y|x) \\
\label{eq7}
q_t(y^n)&\triangleq&\prod_{i=1}^n q_t(y_i) \\
\label{8}
I(t;P)&\triangleq&\sum_{x\in\mathcal{X}}t(x)\sum_{y\in\mathcal{Y}}p(y|x)\ln\frac{p(y|x)}{q_t(y)} 
\end{eqnarray}
where $y^n=y_1y_2\ldots y_n$, and $P=\{p(y|x)\}$ represents the conditional probability between $X$ and $Y$. Clearly, $D(t,x)$ is the divergence or relative entropy between $p(y|x)$ and $q_t(y)$; and $I(t;P)$ is the mutual information between the source $x^n$ and the side information $y^n$ when the source sequence is distributed according to $t$. To be specific, we denote the pmf of each $X_i$ by $p_X$. Without loss of generality, we assume that $p_X(x)>0$ for any $x\in\mathcal{X}$.
Define for any $t\in\mathcal{P}$ with full support and any $\delta\ge0$,
\begin{equation}
\begin{aligned}
&r_-(t,\delta) \\
\label{eq9}
&\triangleq\sup_{\lambda\ge0}\bigg[\lambda(\delta-I(t;P))-\sum_{x\in\mathcal{X}}t(x)\ln\sum_{y\in\mathcal{Y}}p(y|x)\bigg[\frac{p(y|x)}{q_t(y)}\bigg]^{-\lambda}\bigg] 
\end{aligned}
\end{equation}
Define
\begin{equation}
\label{eq10}
\lambda^*_-(X;Y)\triangleq\sup\bigg\{\lambda\ge0:\sum_{x\in\mathcal{X}}\sum_{y\in\mathcal{Y}}p(x,y)\bigg[\frac{p(y|x)}{p(y)}\bigg]^{-\lambda}<\infty\bigg\}
\end{equation}
and for any $\lambda\in[0,\lambda^*_-(X;Y))$ and $t\in\mathcal{P}$ with full support
\begin{equation}
\label{eq11}
f_{-\lambda}(y|x)\triangleq\frac{\left[\frac{p(y|x)}{q_t(y)}\right]^{-\lambda}}{\sum_{v\in\mathcal{Y}}\left[\frac{p(v|x)}{q_t(v)}\right]^{-\lambda}}
\end{equation}
Further define for any $\lambda\in[0,\lambda^*_-(X;Y))$
\begin{equation}
\begin{aligned}
&\sigma^2_{D,-}(t;P,\lambda) \\
\label{eq12}
&\triangleq\sum_{x\in\mathcal{X}}t(x)\left[\sum_{y\in\mathcal{Y}}p(y|x)f_{-\lambda}(y|x)\bigg|\ln\frac{p(y|x)}{q_t(y)}-D(t,x,\lambda)\bigg|^2\right] 
\end{aligned}
\end{equation}
and
\begin{equation}
\begin{aligned}
&M_{D,-}(t;P,\lambda) \\
\label{eq13}
&\triangleq\sum_{x\in\mathcal{X}}t(x)\left[\sum_{y\in\mathcal{Y}}p(y|x)f_{-\lambda}(y|x)\bigg|\ln\frac{p(y|x)}{q_t(y)}-D(t,x,\lambda)\bigg|^3\right]
\end{aligned}
\end{equation}
It is not hard to see that
\begin{equation}
\begin{aligned}
&\sigma^2_D(t;P) \\
\label{eq14}
&=\sum_{x\in\mathcal{X}}t(x)\left[\sum_{y\in\mathcal{Y}}p(y|x)\bigg|\ln\frac{p(y|x)}{q_t(y)}\bigg|^2-\bigg(\sum_{v\in\mathcal{Y}}\ln\frac{p(v|x)}{q_t(v)}\bigg)^2\right]
\end{aligned}
\end{equation}
\begin{equation}
\begin{aligned}
&\sigma^2_D(X;Y) \\
\label{eq15}
&=\sum_{x\in\mathcal{X}}p(x)\left[\sum_{y\in\mathcal{Y}}p(y|x)\bigg|\ln\frac{p(y|x)}{p(y)}\bigg|^2-\bigg(\sum_{v\in\mathcal{Y}}\ln\frac{p(v|x)}{p(v)}\bigg)^2\right]
\end{aligned}
\end{equation}
{\bfseries Theorem 2:} For any sequence $x^n=x_1x_2\ldots x_n$ from $\mathcal{X}$, let $t\in\mathcal{P}$ be the type of $x^n$. Assume that $t$ has full support. Then 
\begin{equation}
\label{eq16}
\Pr\bigg\{\frac{1}{n}\ln\frac{p(Y^n|X^n)}{q_t(Y^n)}\le I(t;P)-\delta\bigg|X^n=x^n\bigg\}\le e^{-nr_-(t,\delta)}
\end{equation}
Furthermore, under the assumptions $\lambda^*_-(X|Y)>0$ and $\sigma^2_D(X;Y)>0$, the following also holds,
\begin{enumerate}
\item There exists a $\delta^*>0$ such that for any $\delta\in(0,\delta^*]$
\begin{equation}
\label{eq17}
r_-(t,\delta)=\frac{1}{2\sigma^2_D(t;P)}\delta^2+O(\delta^3)
\end{equation}
and hence
\begin{eqnarray}
\label{eq18}
&&\Pr\bigg\{\frac{1}{n}\ln\frac{p(Y^n|X^n)}{q_t(Y^n)}\le I(t;P)-\delta\bigg|X^n=x^n\bigg\} \nonumber \\
&\le&e^{-n(\frac{\delta^2}{2\sigma^2_D(t;P)}+O(\delta^3))}
\end{eqnarray}
\item For any $\delta\in(0,\Delta^*_-(X;Y))$
\begin{eqnarray}
\label{eq19}
&&\Pr\bigg\{\frac{1}{n}\ln\frac{p(Y^n|X^n)}{q_t(Y^n)}\le I(t;P)-\delta\bigg|X^n=x^n\bigg\} \nonumber \\
&\ge&\underline{\xi}_{D,-}(t;P,\lambda,n)e^{-nr_-(t,\delta)}
\end{eqnarray}
where $\lambda=\frac{\partial r_-(t,\delta)}{\partial \delta}>0$, and
\begin{equation}
\label{eq20}
\underline{\xi}_{D,-}(t;P,\lambda,n)=e^{\frac{n\lambda^2\sigma^2_{D,-}(t;P,\lambda)}{2}}Q(\rho_*+\sqrt{n}\lambda\sigma_{D,-}(t;P,\lambda))
\end{equation}
with $Q(\rho_*)=\frac{1}{2}-\frac{2CM_{D,-}(t;P,\lambda)}{\sqrt{n}\sigma^3_{D,-}(t;P,\lambda)}$.
\end{enumerate}
%
\section{Achievability}
In the beginning, we need to formally define the variable rate SW coding. The set of binary codewords satisfying the prefix condition is denoted as $\mathcal{I}$. An order $n$ SW code $C_n$ is described by the encoder $f_n(.): \mathcal{X}^n\to\mathcal{I}$ and the decoder $g_n(.,.): \mathcal{I}\times\mathcal{Y}^n\to\mathcal{X}^n$, where $f_n$ maps the source sequence of block length $n$ from $\mathcal{X}$ to the binary codeword in $\mathcal{I}$, and $g_n$ reproduces the source sequence from the received codeword with the help of the side information. Here, we regard a codeword $b\in\mathcal{I}$ as the index of a bin which consists of all the source sequences $x^n$ satisfying $f_n(x^n)=b$. If all the codewords in $\mathcal{I}$ are of the same length, the SW code $C_n$ is regarded as the fixed rate code. \\
The output of the decoder is denoted as $\hat{X}^n=g_n(f_n(X^n), Y^n)$. The decoding error probability of $C_n$ is defined as
\begin{equation}
P_e(C_n)\triangleq\Pr\{X^n\neq\hat{X}^n\} \nonumber
\end{equation}
The average coding rate of $C_n$ is defined as
\begin{equation}
r(C_n)\triangleq\frac{1}{n}{\pmb E}|f_n(C_n)| \nonumber 
\end{equation}
In the rest of the paper, we will investigate the achievability and converse of the redundancy of SW coding, where the decoding error probability is bounded by a given $\epsilon_n$. That reveals a trade-off between the error probability and the redundancy.
\subsection{Variable Rate}
{\bfseries Theorem 3:} Assume $P_{XY} \in \mathcal{P}^{+}(\mathcal{X}\times\mathcal{Y})$ and $I(X;Y)>0$. Let $\{\epsilon_{n}\}$ be a sequence of positive real numbers such that $-\ln\epsilon_{n}=o(n)$ and $\lim_{n\to\infty}\epsilon_{n}=0$. Then there exists a sequence of {\em variable rate} codes $\{C_{n}\}^{\infty}_{n=1}$ with $P_{e}(C_{n})\le\epsilon_{n}$ such that for sufficiently large $n$
\begin{align}
\label{eq21}
r(C_{n})\le&H(X|Y)+\sigma_{D}(P_X;P)\sqrt{\frac{-\ln\epsilon_{n}}{n}} \nonumber \\
&+O\bigg(\frac{1}{\sqrt{n}}\bigg)+O\left(\frac{\ln{n}}{n}\right)+O\left(\frac{-\ln{\epsilon_n}}{n}\right)
\end{align}
where $P=P_{Y|X}$.\\
{\em Proof:} To prove this theorem, we can construct a sequence of codes $\{C_{n}\}^{\infty}_{n=1}$ with desired decoding error probability and redundancy.  \\
Define 
\[\Gamma_X=\left\{t\in\mathcal{T}_n(\mathcal{X}):||t-P_X||_1\le{c_0}\sqrt{\frac{\ln{n}}{n}}\right\} \]
where $c_0$ is selected such that $\Pr\{\tau(X^n)\not\in\Gamma_{\mathcal{X}}\}\le\frac{1}{n^2}$.\\
For any given $t\in\mathcal{T}_{n}(\mathcal{X})$, and $x^{n}$ with type $t$, the encoder $f_{n}$ encodes $x^{n}$ in the following way.
{\begin{quote} Step 1: Encode $t$ by using
\begin{eqnarray}
\label{eq22}
\ln |\mathcal{T}_{n}(\mathcal{X})|&=&\ln {n+|\mathcal{X}|-1 \choose |\mathcal{X}|-1} \nonumber \\
&=&O(\ln{n}) \mbox{bits.}
\end{eqnarray}
Step 2: If $||t-P_{X}||\ge c_{0}\sqrt{\ln{n}/n}$, encode $x^n$ losslessly by using
\begin{equation}
\label{eq23}
\ln|T^n_{\mathcal{X}}(t)| \le [nH(t)] \mbox{bits.}
\end{equation}
Otherwise, throw all $x^n$ with type $t$ randomly into $e^{nR(t)}$ bins, and encode the bin index by $nR(t)$ bits, where $R(t)$ will be specified later. Note that the random bin partition is independent of $(X,Y)$, and is known to both encoder and decoder. The codeword length $|f_n(x^n)|$ only depends on the type $t$ of $x^n$.
\end{quote}}
On the decoder side, define the jar centered at side information sequence $y^n$ by
\begin{equation}
\label{eq24}
J_t(y^n) = \left\{x^n\in\mathcal{X}^n : \frac{1}{n}\ln\frac{p(y^n|x^n)}{q_t(y^n)}\ge I(t;P)-\delta_n\right\}
\end{equation}
and select
\begin{eqnarray}
R(t)=H(t\circ P_{Y|X})-H((t\circ P_{Y|X})_{\mathcal{Y}})+\delta_n \nonumber \\
\label{eq25}
+\kappa_2\frac{-\ln\epsilon_n}{n}+O\bigg(\frac{-\ln\epsilon_n}{n}\bigg)
\end{eqnarray}
where $\delta_n=\sigma_D(t;P)\sqrt{\frac{-\ln\epsilon_n/\kappa_1}{n}}$, with $\kappa_1>1$ and $\kappa_2>0$. \\
The decoder mapping works in the following way.
\begin{quote}
Step 1: Decode t from the transmitted codeword.
Step 2: If $||t-P_X||_1 \ge c_0\sqrt{\ln{n}/n}$, decode the source sequence $x^n$ from the transmitted codeword; otherwise, decode the bin index from the received sequence, and continue to Step 3 below. \\
Step 3: If $||t-P_X||_1 \le c_0\sqrt{\ln{n}/n}$. For any $x^n$ with type $t\in\mathcal{T}_n(\mathcal{X})$, decode the bin index $i^*$ from the received sequence. With any side information sequence $y^n\in\mathcal{Y}^n$, we can find $\hat{x}^n\in\mbox{BIN}_{i^*}\cap J_t(y^n)$. Denote $\hat{x}^n$ as the estimate of the source sequence. We can prove that with high probability, there is only one source sequence $x^n$ lying in $\mbox{BIN}_{i^*}\cap J_t(y^n)$. If $\mbox{BIN}_{i^*}\cap J_t(y^n)$ contains more than one sequence, the decoder selects $\hat{x}^n$ arbitrarily.
\end{quote}
{\em Analysis of error probability:} Suppose that $\tau(X^n)=t$. If $||t-P_X||_1\ge c_0\sqrt{\ln{n}/n}$, the decoder decodes the source sequence $x^n$ from the transmitted codeword, where the error probability is zero. If $||t-P_X||_1<c_0\sqrt{\ln{n}/n}$, we see that a decoding error happens if either one of the following 
two events occurs. 
\begin{quote}
i)  $X^n\notin J_t(Y^n)$\\
ii) There exists another different sequence $\hat{X}^n\in\mbox{BIN}_{i^*}\cap J_t(Y^n)$.
\end{quote}
In view of these, we have
\begin{eqnarray}
&&\Pr\{\hat{X}^n\neq X^n|\tau(X^n)=t\} \nonumber \\
&\le&\Pr\{X^n\not\in J_t(Y^n)|\tau(X^n)=t\} \nonumber \\
\label{eq26}
&&+\Pr\{\hat{X}^n\neq X^n, \hat{X}^n\in\mbox{BIN}_{i^*}\cap J_t(Y^n)|\tau(X^n)=t\}  
\end{eqnarray}
Next, our job is to upper bound the two terms on the right hand side of (34). 

To upper bound $\Pr\{Y^n\not\in J_t(X^n)|\tau(X^n)=t\}$, we apply (25),
\begin{eqnarray}
&&\Pr\{Y^n\not\in J_t(X^n)|\tau(X^n)=t\} \nonumber \\
&\le&\Pr\left\{\frac{1}{n}\sum_{i=1}^n\ln\frac{p(Y_i|X_i)}{q_t(Y_i)}<I(t;P)-\delta_n|\tau(X^n)=t\right\} \nonumber \\
&\le&e^{-n(\frac{\delta_n^2}{2\sigma^2_D(t;P)}+O(\delta_n^3))} \nonumber \\
\label{eq27}
&\le&e^{-n(\frac{-\ln{\epsilon_n}/\kappa_1}{n}+O((\frac{-\ln{\epsilon_n}}{n})^{\frac{3}{2}}))}
\end{eqnarray}
So, with proper choice of $\kappa_1$, the probability above can be less than $\epsilon_n/2$.
From the definition of $J_t(y^n)$, we can conclude,
\begin{eqnarray}
\frac{p(y^n|x^n)}{q_t(y^n)}&=&\frac{p(x^n, y^n)}{q_t(y^n)p(x^n)} \nonumber \\
&=&\frac{p(x^n|y^n)}{p(x^n)} \nonumber \\
\label{eq28}
&\ge&e^{n(I(t;P)-\delta_n)}
\end{eqnarray}
From the property of type in \cite{type}, we have $p(x^n)=e^{-nH(t)}$. Hence, we have
\begin{equation}
\label{eq29}
p(x^n|y^n)\ge e^{-nH(t)}e^{n(I(t;P)-\delta_n)}
\end{equation}
Since $\sum{p(x^n|y^n)}=1$, with fixed $y^n$, we can derive,
\begin{equation}
\label{eq30}
|\{x^n\in T^n_{\mathcal{X}}(t):y^n\in J_t(x^n)\}|\le e^{(n(H(t)-I(t;P)+\delta_n))}
\end{equation}
Hence, the second term in \eqref{eq7} can be upper bounded as below,
\begin{eqnarray}
&&Pr\{\hat{X}^n\neq X^n, \hat{X}^n\in\mbox{BIN}_{i^*}\cap J_t(Y^n)|\tau(X^n)=t\} \nonumber \\
&\le&\sum_{\hat{X}^n:\hat{X}^n\in J_t(Y^n), \tau(\hat{X}^n)=t}{\Pr\{\hat{X}^n\in\mbox{BIN}_i|\tau(\hat{X}^n)=t\}} \nonumber \\
&\le&|\{x^n\in T^n_{\mathcal{X}}(t):x^n\in J_t(y^n)\}|e^{-nR(t)} \nonumber \\
&\le&e^{n(H(t)-R(t)-I(t;P)+\delta_n)} \nonumber \\
\label{eq31}
&\le&e^{-n(\sigma_D(t;P)\sqrt{\frac{-\ln{\epsilon_n}/\kappa_1}{n}}+\kappa_2\frac{-\ln{\epsilon_n}}{n}+O(\frac{-\ln{\epsilon_n}}{n}))}
\end{eqnarray}
Obviously, it is possible to find proper constant $\kappa_2$ to make  smaller than $\epsilon_n/2$. Combining \eqref{eq27} and \eqref{eq31}, \eqref{eq26} can be smaller than $\epsilon_n$.

Next, we need to average out the type $t$. From \cite{he}, we know that $\sigma_D(t;P)$ is concave with respect to $t$. So, 
\begin{equation}
\label{eq32}
\sigma_D(t;P)\ge\sigma_D(P_X;P)-\kappa_6||t-P_X||_1
\end{equation}
Define $F(t)=H(t\circ{P})-H((t\circ{P})_{\mathcal{Y}})$. Expanding $F(t)$ at $P_X$ by using Taylor's series, we have
\begin{eqnarray}
F(t)&=&F(P_X)+\frac{\partial{F(t)}}{\partial{t}}\bigg|_{t=P_X}(t-P_X)' \nonumber \\
&&+\frac{1}{2}(t-P_X)\frac{\partial^2F(t)}{\partial{t}^2}\bigg|_{t=P_X}(t-P_X)' \nonumber \\
&&+o(||t-P_X||^2_1) \nonumber \\
&=&H(X|Y)+\frac{\partial{F(t)}}{\partial{t}}\bigg|_{t=P_X}(t-P_X)' \nonumber \\
&&+\frac{1}{2}(t-P_X)\frac{\partial^2F(t)}{\partial{t}^2}\bigg|_{t=P_X}(t-P_X)' \nonumber \\
\label{eq33}
&&+o(||t-P_X||^2_1) 
\end{eqnarray}
Let us introduce two properties of the type $t$. First, it is obvious that
\begin{equation}
\label{eq34}
\sum_{t\in\mathcal{T}_n(\mathcal{X})}p(t)(t-P_X)=0
\end{equation}
Second, from \cite{he}, we have
\begin{equation}
\label{eq35}
\sum_{t\in\mathcal{T}_t(\mathcal{X})}p(t)\frac{1}{2}(t-P_X)\frac{\partial^2F(t)}{\partial{t}^2}\bigg|_{t=P_X}(t-P_X)'=O\left(\frac{1}{n}\right)
\end{equation}
There is another useful inequality to be applied in the following derivation. For $x>0$ and $y>0$, we have
\begin{equation}
\label{eq36}
\sqrt{x+y}<\sqrt{x}+\sqrt{y}
\end{equation}
So, together with \eqref{eq33}-\eqref{eq36}, based on the encoding scheme above, the rate of our code $C_n$ can be bounded as,
\begin{eqnarray}
&&r(C_n) \nonumber \\
&\le&\sum_{t\in\Gamma_{\mathcal{X}}}Pr\{\tau(X^n)=t\}R(t)+\sum_{t\not\in\Gamma_{\mathcal{X}}}Pr\{\tau(X^n)=t\}H(t) \nonumber \\
&&+O\left(\frac{\ln{n}}{n}\right) \nonumber \\
&\le&\sum_{t\in\Gamma_{\mathcal{X}}}Pr\{\tau(X^n)=t\}\bigg[H(t\circ P_{Y|X})-H((t\circ P_{Y|X})_{\mathcal{Y}}) \nonumber \\
&&+\sigma_D(t;P)\sqrt{\frac{-\ln\epsilon_n/\kappa_1}{n}}+O\bigg(\frac{-\ln\epsilon_n}{n}\bigg)\bigg] \nonumber  \\
&&+\frac{\ln{|\mathcal{X}|}}{n^2}+O\left(\frac{\ln{n}}{n}\right) \nonumber \\
&\le&H(X|Y)+\sigma_D(P_X;P)\sqrt{\frac{-\ln\epsilon_n+\ln\kappa_1}{n}} \nonumber \\
&&+\sum_{t\in\Gamma_{\mathcal{X}}}p(t)\frac{\partial{F(t)}}{\partial{t}}\bigg|_{t=P_X}(t-P_X)' \nonumber \\
&&+\sum_{t\in\Gamma_{\mathcal{X}}}\frac{1}{2}(t-P_X)\frac{\partial^2F(t)}{\partial{t}^2}\bigg|_{t=P_X}(t-P_X)' \nonumber \\
&&+\frac{\ln|\mathcal{X}|}{n^2}+O\left(\frac{\ln{n}}{n}\right)+O\bigg(\frac{-\ln\epsilon_n}{n}\bigg) \nonumber
\end{eqnarray}
\begin{eqnarray}
&\le&H(X|Y)+\sigma_D(P_X;P)\bigg(\sqrt{\frac{-\ln{\epsilon_n}}{n}}+\sqrt{\frac{\ln\kappa_1}{n}}\bigg) \nonumber \\
&&+O\bigg(\frac{1}{n}\bigg)+O\left(\frac{\ln{n}}{n}\right)+O\left(\frac{-\ln{\epsilon_n}}{n}\right) \nonumber \\
&=&H(X|Y)+\sigma_D(P_X;P)\sqrt{\frac{-\ln{\epsilon_n}}{n}}+O\bigg(\frac{1}{\sqrt{n}}\bigg) \nonumber \\
\label{eq37}
&&+O\bigg(\frac{1}{n}\bigg)+O\left(\frac{\ln{n}}{n}\right)+O\left(\frac{-\ln{\epsilon_n}}{n}\right)
\end{eqnarray}
\subsection{Fixed Rate}
{\bfseries Theorem 4:} Assume $P_{XY}\in\mathcal{P}^+(\mathcal{X}\times\mathcal{Y})$. Further assume either $I(X;Y)>0$ or $X$ is not uniformly distributed. Let $\{\epsilon_n\}$ be a sequence of positive real numbers such that $-\ln\epsilon_n=o(n)$ and $\lim_{n\to\infty}{\epsilon_n}=0$. Then there exists a sequence of {\em fixed rate} codes $\{C_n\}^{\infty}_{n=1}$ with $P_e(C_n)<\epsilon_n$ such that for sufficiently large $n$,
\begin{eqnarray}
\label{eq38}
r(C_n)\le H(X|Y)+\sigma_H(X|Y)\sqrt{\frac{-\ln\epsilon_n}{n}} \nonumber \\
+O\bigg(\frac{1}{\sqrt{n}}\bigg)+O\bigg(\frac{-\ln\epsilon_n}{n}\bigg)
\end{eqnarray}

{\em Proof:} Fixed rate coding is a special case of variable rate coding. For any side information sequence $y^n\in\mathcal{Y}^n$, we define a jar as,
\begin{equation}
\label{eq39}
J(y^n) = \left\{x^n\in\mathcal{X}^n:-\frac{1}{n}\ln{p(x^n|y^n)}\ge H(X|Y)-\delta_n\right\}
\end{equation}
where $\delta_n=\sigma_H(X|Y)\sqrt{\frac{-\ln\epsilon_n/\kappa_3}{n}}$.
First, we select the rate $R$ as below,
\begin{equation}
\label{eq40}
R=H(X|Y)+\delta_n+\kappa_4\frac{-\ln\epsilon_n}{n}+O\left(\frac{-\ln\epsilon_n}{n}\right)
\end{equation}
where $\kappa_3>1$ and $\kappa_4>0$. \\
{\em Codebook generation:} Throw all the source sequence $x^n$ randomly into $e^{nR}$ bins. And the bin partition is known to both encoder and decoder.\\
{\em Encoding:} Transmit the index of the bin containing $x^n$, namely $f_n(x^n)$, to the decoder. \\
{\em Decoding:} Decode the bin index $i^*$ from the received sequence. For any fixed side information $y^n$, find $\hat{x}^n\in\mbox{BIN}_{i^*}\cap J(y^n)$. We can show that with high probability, there exists one and only one such $\hat{x}^n$ in $J(y^n)\cap\mbox{BIN}_{i^*}$.

We see a decoding error happens if one of the following two events occurs.

i) $X^n\not\in J(Y^n)$,

ii)$X^n\neq\hat{X}^n$ and $\hat{X}^n\in J(y^n)\cap\mbox{BIN}_{i^*}$.\\
So, we have,
\begin{eqnarray}
P_e&=&\Pr\{X^n\neq\hat{X}^n\} \nonumber \\
&\le&\Pr\{X^n\not\in J(Y^n)\} \nonumber \\
\label{eq41}
&&+\Pr\{X^n\neq\hat{X}^n, \hat{X}^n\in J(y^n)\cap\mbox{BIN}_{i^*}\}
\end{eqnarray}
The first term in \eqref{eq41} can be bounded by \eqref{eq5},
\begin{eqnarray}
\Pr\{X^n\not\in J(Y^n)\}&\le&e^{-n(\frac{\delta_n^2}{2\sigma^2_H(X|Y)}+O(\delta_n^3))} \nonumber \\
\label{eq42}
&\le&e^{-n({\frac{-\ln\epsilon_n/\kappa_3}{n}}+O\left(\left(\frac{-\ln\epsilon_n}{n}\right)^{\frac{3}{2}}\right)}
\end{eqnarray}
So, there exists a proper constant $\kappa_3$ which can make \eqref{eq42} smaller than $\epsilon_n/2$.
We can use the similar technique in proof of Theorem 3 to bound the second term in \eqref{eq41}.
\begin{eqnarray}
&&\Pr\{X^n\neq\hat{X}^n, \hat{X}^n\in J(Y^n)\cap\mbox{BIN}_{i^*}\} \nonumber \\
&=&\Pr\{X^n\neq\hat{X}^n, \hat{X}^n\in J(Y^n)\cap\mbox{BIN}_{i^*}|X^n=x^n\} \nonumber \\
&\le&\sum_{\hat{X}^n\in J(Y^n): \hat{X}^n\neq X^n}{\Pr\{f_n(X^n)=f_n(\hat{X}^n)\}} \nonumber \\
&\le&e^{nH(X|Y)}e^{-nR} \nonumber \\
\label{eq43}
&\le&e^{-n\left(\sigma_H(X|Y)\sqrt{\frac{-\ln\epsilon_n/\kappa_3}{n}}+\kappa_4\frac{-\ln\epsilon_n}{n}+O\left(\frac{-\ln\epsilon_n}{n}\right)\right)}
\end{eqnarray}
With proper choice of $\kappa_4$, \eqref{eq43} can be bounded by $\epsilon_n/2$.

Together with \eqref{eq42} and \eqref{eq43}, we can conclude that when the rate $R$ is selected as \eqref{eq38}, the error probability $P_e$ can be bounded by $\epsilon_n$.
So, there exists a code $C_n$, such that for $n$ large enough, with $P_e(C_n)<\epsilon_n$, 
\begin{eqnarray}
r(C_n)&\le&H(X|Y)+\sigma_H(X|Y)\sqrt{\frac{-\ln\epsilon_n}{n}} \nonumber \\
\label{eq44}
&&+O\bigg(\frac{1}{\sqrt{n}}\bigg)+O\bigg(\frac{-\ln\epsilon_n}{n}\bigg)
\end{eqnarray}


\section{Converse}
\subsection{Variable Rate}
{\bfseries Theorem 5:} Assume $P_{XY}\in\mathcal{P}^+(\mathcal{X}\times\mathcal{Y})$ and $I(X;Y)>0$. Let $\{\epsilon_n\}$ be a sequence of positive real numbers satisfying $\epsilon_n=\frac{1}{\sqrt{n\ln{n}}}$. Then for sufficiently large $n$ and any order $n$ {\em variable rate} code $C_n=(f_n,g_n)$ with
\begin{equation}
\label{eq45}
P_e(C_n)=\Pr\{X^n\neq\hat{X}^n\}\le\epsilon_n
\end{equation}
one has
\begin{eqnarray}
r(C_n)\ge H(X|Y)+\sigma_D(P_X;P)\sqrt{\frac{-\ln\epsilon_n}{n}} \nonumber \\
\label{eq46}
+O\left(\frac{-\ln\epsilon_n}{n}\right)-O\left(\frac{\ln{n}}{n}\right)-O\bigg(\frac{1}{\sqrt{n\ln{n}}}\bigg)
\end{eqnarray}
{\em Proof:} Let $C_n=(f_n, g_n)$ be an order $n$ variable rate code with
\begin{eqnarray}
P_e(C_n)&=&\Pr\{X^n\neq\hat{X}^n\} \nonumber \\
\label{eq47}
&=&\sum_{x^n\in\mathcal{X}^n}{\Pr\{X^n=x^n\}\epsilon_{x^n}}\le\epsilon_n
\end{eqnarray}
where $\epsilon_{x^n}\triangleq\Pr\{\hat{X}^n\neq X^n|X^n=x^n\}$. We define $b\in\mathcal{I}$ as the codeword for transmission, where $\mathcal{I}$ is a prefix set. For any $b\in\mathcal{I}$ and $t\in\mathcal{T}_n(\mathcal{X})$, define
\begin{eqnarray}
p(t)&\triangleq&\Pr\{\tau(X^n)=t\} \nonumber \\
p(b,t)&\triangleq&\Pr\{f_n(X^n)=b, \tau(X^n)=t\} \nonumber \\
p(b|t)&\triangleq&\Pr\{f_n(X^n)=b|\tau(X^n)=t\} \nonumber
\end{eqnarray}
Then, we can easily see that
\begin{eqnarray}
nr(C_n)&\ge&H(f_n(X^n))\ge H(f_n(X^n)|\tau(X^n)) \nonumber \\
\label{eq49}
&=&\sum_{t\in\mathcal{T}_n(\mathcal{X})}{p(t)H(f_n(X^n)|\tau(X^n)=t)}
\end{eqnarray}
Therefore, we shall consider only the case for some certain $t\in\mathcal{T}_n(\mathcal{X})$. With the definition of entropy, we have for any $t\in\mathcal{T}_n(\mathcal{X})$,
\[
H(f_n(X^n)|\tau(X^n)=t)=\sum_{b\in\mathcal{I}}{p(b|t)\ln\frac{p(t)}{p(b,t)}} \]
which together with \eqref{eq49} implies,
\begin{equation}
\label{eq50}
nr(C_n)\ge\sum_{t\in\mathcal{T}_n(\mathcal{X})}p(t)\left[\sum_{b\in\mathcal{I}}{p(b|t)\ln\frac{p(t)}{p(b,t)}}\right]
\end{equation}
Since every sequence $x^n$ in $\mathcal{T}^n_{\mathcal{X}}(t)$ is equally probable, we have
\begin{equation}
\label{eq51}
\frac{p(t)}{p(b,t)}=\frac{|\{x^n\in\mathcal{X}^n:\tau(x^n)=t\}|}{|\{x^n:f_n(x^n)=b,\tau(x^n)=t\}|}
\end{equation}
Define
\begin{eqnarray}
\mathcal{N}_{b,t,n}&\triangleq&\{x^n:f_n(x^n)=b,\tau(x^n)=t\} \nonumber \\
\epsilon_{b,t,n}&\triangleq&\Pr\{\hat{X}^n\neq X^n|f_n(X^n)=b,\tau(X^n)=t\} \nonumber 
\end{eqnarray}
Thus,
\begin{eqnarray}
\epsilon_{b,t,n}&=&\sum_{x^n\in\mathcal{N}_{b,t,n}}{\frac{\epsilon_{x^n}}{|\mathcal{N}_{b,t,n}|}} \nonumber \\
\epsilon_n&=&\sum_{t\in\mathcal{T}_n(\mathcal{X})}\sum_{b\in\mathcal{I}}p(b,t)\epsilon_{b,t,n} \nonumber
\end{eqnarray}
Based on the duality of channel coding and Slepian Wolf coding, $\mathcal{N}_{b,t,n}$ can be regarded as the set of channel codes. This channel is discrete memoryless, defined by $P_{Y|X}$. We can use the concept of jar to upper bound $|\mathcal{N}_{b,t,n}|$. \\
First, define 
\begin{eqnarray}
B_t(x^n,\delta_n)&\triangleq&\left\{y^n:\frac{1}{n}\ln\frac{p(y^n|x^n)}{q_t(y^n)}<I(t;P)-\delta_n\right\} \nonumber \\
B_{t,\delta_n}&\triangleq&\cup_{x^n\in\mathcal{N}_{b,t,n}}B_t(x^n,\delta_n) \nonumber \\
\mathcal{M}_{t}&\triangleq&\{x^n\in\mathcal{N}_{b,t,n}:\epsilon_{x^n}\le\epsilon_{n}(1+\beta_n)\} \nonumber
\end{eqnarray}
where $\beta_n>0$ will be specified later. By Markov inequality,
\begin{equation}
\label{eq52}
\Pr\{x^n\in\mathcal{M}_{t}|x^n\in\mathcal{N}_{b,t,n}\}=\frac{|\mathcal{M}_{t}|}{|\mathcal{N}_{b,t,n}|}\ge\frac{1}{1+\beta_n}
\end{equation}
Denote the decision region for message $x^n\in\mathcal{M}_{t}$ as $D_{x^n}$. Therefore,
\begin{eqnarray}
&&P_{x^n}(B_t(x^n,\delta_n)\cap D_{x^n}) \nonumber \\
&=&P_{x^n}(B_t(x^n,\delta_n))-P_{x^n}(B_t(x^n,\delta_n)\cap D^c_{x^n}) \nonumber \\
&\ge&P_{x^n}(B_t(x^n,\delta_n))-\epsilon_{x^n} \nonumber \\
\label{eq53}
&\ge&P_{x^n}(B_t(x^n,\delta_n))-\epsilon_{n}(1+\beta_n)
\end{eqnarray}
We can select $\delta_n$ such that for any $x^n$,
\begin{equation}
\label{eq54}
P_{x^n}(B_t(x^n,\delta_n))=P_{0^n}(B_t(0^n,\delta_n))\ge\epsilon_{n}(1+2\beta_n)
\end{equation}
Plugging \eqref{eq54} into \eqref{eq53} yields
\begin{equation}
\label{eq55}
P_{x^n}(B_t(x^n,\delta_n)\cap D_{x^n})\ge\beta_n\epsilon_{n}
\end{equation}
By the fact that $D_{x^n}$ are disjoint for different $x^n$ and
\begin{equation}
\label{eq56}
\cup_{x^n}(B_t(x^n,\delta_n)\cap D_{x^n})\subseteq B_{t,\delta_n}
\end{equation}
we have
\begin{eqnarray}
&&P(B_{t,\delta_n}) \nonumber \\
&=&\sum_{y^n\in B_{t,\delta_n}}p(y^n) \nonumber \\
&\ge&\sum_{x^n\in\mathcal{N}_{b,t,n}}\sum_{y^n\in B_t(x^n,\delta_n)\cap D_{x^n}}p(y^n) \nonumber \\
&\ge&\sum_{x^n\in\mathcal{M}_t}\sum_{y^n\in B_t(x^n,\delta_n)\cap D_{x^n}}q_t(y^n) \nonumber \\
&\ge&\sum_{x^n\in\mathcal{M}_t}\sum_{y^n\in B_t(x^n,\delta_n)\cap D_{x^n}}p(y^n|x^n)e^{n(-I(t;P)+\delta_n)} \nonumber \\
&=&\sum_{x^n\in\mathcal{M}_t}e^{n(-I(t;P)+\delta_n)}P_{x^n}(B_t(x^n,\delta_n)\cap D_{x^n}) \nonumber \\
&\ge&\sum_{x^n\in\mathcal{M}_t}e^{n(-I(t;P)+\delta_n)}\beta_n\epsilon_{n} \nonumber \\
\label{eq57}
&=&|\mathcal{M}_t|e^{n(-I(t;P)+\delta_n)}\beta_n\epsilon_{n}
\end{eqnarray}
which implies that
\begin{equation}
\label{eq58}
|\mathcal{M}_t|\le e^{n(I(t;P)-\delta_n-\ln\beta_n-\ln\epsilon_{n}+\ln{P(B_{t,\delta_n})})}
\end{equation}
Combining \eqref{eq52} and \eqref{eq58} yields
\begin{equation}
\label{eq59}
|\mathcal{N}_{b,t,n}|\le e^{n(I(t;P)-\delta_n-\frac{\ln\frac{\beta_n}{1+\beta_n}}{n}-\frac{\ln\epsilon_{n}-\ln P(B_{t,\delta_n})}{n})}
\end{equation}
Define $\beta_n=1$. So, $\delta_n$ should satisfy
\begin{equation}
\label{eq60}
P_{x^n}(B_t(x^n,\delta_n))\ge3\epsilon_{n}
\end{equation}
where $\epsilon_n=\frac{n^{-\alpha}}{\sqrt{\ln{n}}}$ for $\alpha>0$.
Assume $\delta_n=\sigma_D(t;P)\sqrt{\frac{2\alpha\ln{n}}{n}}-\eta\sqrt{\frac{1}{n\ln{n}}}$. Then, in order to satisfy \eqref{eq60}, we will show that with some properly chosen constant $\eta$, 
\begin{eqnarray}
&&P_{x^n}(B_t(x^n,\delta_n)) \nonumber \\
&\ge&\underline{\xi}_{D,-}\bigg(t;P,\frac{\partial{r_-(t,\delta)}}{\partial\delta},n\bigg)e^{-nr_-(t,\delta_n)} \nonumber \\
\label{eq61}
&\ge&\frac{3n^{-\alpha}}{\sqrt{\ln{n}}}
\end{eqnarray}
From \eqref{eq17} and \eqref{eq18}, we have
\begin{eqnarray}
&&e^{-nr_-(t,\delta_n)} \nonumber \\
&=&e^{-nr_-\left(t,\sigma_D(t;P)\sqrt{\frac{2\alpha\ln{n}}{n}}-\eta\sqrt{\frac{1}{n\ln{n}}}\right)} \nonumber \\
&=&e^{-n\left[\frac{1}{2\sigma_D^2(t;P)}\left(\sigma_D(t;P)\sqrt{\frac{2\alpha\ln{n}}{n}}-\eta\sqrt{\frac{1}{n\ln{n}}}\right)^2+O\left(\sqrt{\frac{\ln^3{n}}{n^3}}\right)\right]} \nonumber \\
&=&e^{-\alpha\ln{n}+\frac{\sqrt{2\alpha}}{\sigma_D(t;P)}+O\left(\frac{1}{\ln{n}}\right)} \nonumber \\
\label{eq62}
&=&e^{-\alpha\ln{n}+\frac{\sqrt{2\alpha}}{\sigma_D(t;P)}+o(1)}
\end{eqnarray}
In addition, based on \eqref{eq20}, there is
\begin{eqnarray}
&&\underline{\xi}_{D,-}\left(t;P,\lambda,n\right) \nonumber \\
&=&e^{\frac{n\lambda^2\sigma_{D,-}^2(t;P,\lambda)}{2}}Q(\sqrt{n}\lambda\sigma_{D,-}(t;P,\lambda))(1-o(1)) \nonumber \\
&=&\Theta\left(\frac{1}{\sqrt{n}\lambda}\right) \nonumber \\
&=&\Theta\left(\frac{1}{\sqrt{\ln{n}}}\right) \nonumber \\
\label{eq63}
&\ge&\frac{\eta_1}{\sqrt{\ln{n}}}
\end{eqnarray}
for some constant $\eta_1>0$, where
\begin{eqnarray}
\lambda=\frac{\partial r_-(t,\delta_n)}{\partial\delta_n}=\frac{\delta_n}{\sigma^2_D(t;P)}+O(\delta^2_n) \nonumber
\end{eqnarray}
Then \eqref{eq61} is satisfied by choosing a constant $\eta$ such that
\begin{equation}
\label{eq64}
e^{\frac{\sqrt{2\alpha}\eta}{\sigma_D(t;P)}+o(1)}\eta_1\ge3
\end{equation}
From the above discussion, substituting $\epsilon_n$ into $\delta_n$, we can derive
\begin{eqnarray}
\delta_n&=&\sigma_D(t;P)\sqrt{\frac{2\alpha\ln{n}}{n}}-\eta\sqrt{\frac{1}{n\ln{n}}} \nonumber \\
\label{eq65}
&=&\sigma_D(t;P)\sqrt{\frac{-2\ln\epsilon_n-\ln\ln{n}}{n}}-\eta\sqrt{\frac{1}{n\ln{n}}}
\end{eqnarray} \\
Choosing $\alpha=0.5$, we have $-\ln\epsilon_n-\ln\ln{n}=\frac{1}{2}\ln\frac{n}{\ln{n}}>0$. Thus,
\begin{equation}
\label{eq66}
-2\ln\epsilon_n-\ln\ln{n}>-\ln\epsilon_n
\end{equation}
Plugging \eqref{eq66} into \eqref{eq65} yields that
\begin{equation}
\label{eq67}
\delta_n\ge\sigma_D(t;P)\sqrt{\frac{-\ln\epsilon_n}{n}}-\eta\sqrt{\frac{1}{n\ln{n}}}
\end{equation}
Based on {\em Lemma II.2} in \cite{type}, it is the fact that $|\mathcal{T}^n_{\mathcal{X}}(t)|=e^{nH(t)-\frac{|\mathcal{X}|-1}{2}\ln{n}+O(1)}$. Together with \eqref{eq51}, \eqref{eq59} and \eqref{eq67}, we have
\begin{eqnarray} 
&&\frac{1}{n}\ln\frac{p(t)}{p(b,t)} \nonumber \\
&=&\frac{1}{n}\ln\frac{|\{x^n:\tau(x^n)=t\}|}{|\mathcal{N}_{b,t,n}|} \nonumber \\
&\ge&H(t\circ P_{Y|X})-H((t\circ P_{Y|X})_{\mathcal{Y}}) \nonumber \\
&&+\sigma_D(t;P)\sqrt{\frac{-\ln\epsilon_{n}}{n}}-\frac{\ln{2}}{n}+\frac{\ln\epsilon_{n}}{n} \nonumber \\\label{eq68}
&&-\frac{|\mathcal{X}|-1}{2}\frac{\ln{n}}{n}+O\left(\frac{1}{n}\right)-O\left(\frac{1}{\sqrt{n\ln{n}}}\right)
\end{eqnarray}
Next, we need to average out $t$ in \eqref{eq68}. There are only two parts of \eqref{eq68} containing $t$. Let us first deal with
\begin{equation}
\label{eq69}
\sum_{t\in\mathcal{T}_n(\mathcal{X})}p(t)(H(t\circ{P})-H((t\circ{P})_{\mathcal{Y}}))
\end{equation}
Define 
\[\Gamma_X=\left\{t\in\mathcal{T}_n(\mathcal{X}):||t-P_X||_1\le{c_0}\sqrt{\frac{\ln{n}}{n}}\right\} \]
where $c_0$ is selected such that $\Pr\{\tau(X^n)\not\in\Gamma_{\mathcal{X}}\}\le\frac{1}{n^2}$.

Together with \eqref{eq33}, \eqref{eq34} and \eqref{eq35}, there is
\begin{eqnarray}
&&\sum_{t\in\mathcal{T}_n(\mathcal{X})}p(t)[H(t\circ{P})-H((t\circ{P})_{\mathcal{Y}})] \nonumber \\
&\ge&\sum_{t\in\Gamma_{\mathcal{X}}}p(t)[H(t\circ{P})-H((t\circ{P})_\mathcal{Y})] \nonumber \\
&=&\sum_{t\in\Gamma_{\mathcal{X}}}p(t)\bigg[H(X|Y)+\frac{\partial{F(t)}}{\partial{t}}\bigg|_{t=P_X}(t-P_X)' \nonumber \\
&&+\frac{1}{2}(t-P_X)\frac{\partial^2F(t)}{\partial{t}^2}\bigg|_{t=P_X}(t-P_X)'\bigg]+o\left(\frac{\ln{n}}{n}\right) \nonumber \\
&\ge&H(X|Y)+\sum_{t\in\mathcal{T}_n(\mathcal{X})}p(t)\bigg[\frac{\partial{F(t)}}{\partial{t}}\bigg|_{t=P_X}(t-P_X)' \nonumber \\
&&+\frac{1}{2}(t-P_X)\frac{\partial^2F(t)}{\partial{t}^2}\bigg|_{t=P_X}(t-P_X)'\bigg]+o\left(\frac{\ln{n}}{n}\right) \nonumber \\
\label{eq70}
&=&H(X|Y)+o\bigg(\frac{\ln{n}}{n}\bigg)
\end{eqnarray}
Invoking {\em Lemma 6} in \cite{he}, we have an important inequality
\begin{equation}
\label{eq71}
\sum_{t\in\mathcal{T}_n(\mathcal{X})}||t-P_X||_1\le\frac{\kappa_5}{\sqrt{n}}
\end{equation}
Based on \eqref{eq32} and \eqref{eq71}, the average of $\sigma_D(t;P)$ over $t$ can be lower bounded as
\begin{eqnarray}
&&\sum_{t\in\mathcal{T}_n(\mathcal{X})}p(t)\sigma_D(t;P)\sqrt{\frac{-\ln\epsilon_{n}}{n}} \nonumber \\
&\ge&\sum_{t\in\mathcal{T}_n(\mathcal{X})}p(t)[\sigma_D(P_X;P)-\kappa_6||t-P_X||_1]\sqrt{\frac{-\ln\epsilon_{n}}{n}} \nonumber \\
\label{eq72}
&\ge&\sigma_D(P_X;P)\sqrt{\frac{-\ln\epsilon_n}{n}}-\frac{\kappa_7}{\sqrt{n}}\sqrt{\frac{-\ln\epsilon_{n}}{n}} 
\end{eqnarray}
Combining \eqref{eq68}, \eqref{eq70} and \eqref{eq72}, we can conclude
\begin{eqnarray}
&&r(C_n) \nonumber \\
&\ge&\sum_{t\in\mathcal{T}_n(\mathcal{X})}\frac{1}{n}p(t)H(f_n(X^n)|\tau(X^n)=t) \nonumber \\
&\ge&H(X|Y)+\sigma_D(P_X;P)\sqrt{\frac{-\ln\epsilon_n}{n}}-\frac{\ln2}{n} \nonumber \\
&&+\frac{\ln\epsilon_n}{n}-\kappa_7\frac{\sqrt{-\ln\epsilon_n}}{n}-\frac{|\mathcal{X}|-1}{2}\frac{\ln{n}}{n} \nonumber \\
&&+o\bigg(\frac{\ln{n}}{n}\bigg)+O\bigg(\frac{1}{n}\bigg)-O\bigg(\frac{1}{\sqrt{n\ln{n}}}\bigg) \nonumber
\end{eqnarray}
\begin{eqnarray}
&=&H(X|Y)+\sigma_D(P_X;P)\sqrt{\frac{-\ln\epsilon_n}{n}} \nonumber \\
\label{eq73}
&&-O\bigg(\frac{\ln{n}}{n}\bigg)+O\bigg(\frac{-\ln\epsilon_n}{n}\bigg)-O\bigg(\frac{1}{\sqrt{n\ln{n}}}\bigg)
\end{eqnarray}


\subsection{Fixed Rate}
{\bfseries Theorem 6:}Assume $P_{XY}\in\mathcal{P}^+(\mathcal{X}\times\mathcal{Y})$. For any order $n$ fixed rate code $C_n=(f_n, g_n)$ with decoding error probability $P_e(C_n)\le\epsilon_n$,
\begin{eqnarray}
r(C_n)&\ge&H(X|Y)+\sigma_H(X|Y)\sqrt{\frac{-\ln\epsilon_n}{n}} \nonumber \\
\label{eq74}
&&-O\bigg(\frac{-\ln\epsilon_n}{n}\bigg)-O\bigg(\frac{1}{\sqrt{n\ln{n}}}\bigg)
\end{eqnarray}
where $\epsilon_n=\frac{1}{\sqrt{n\ln{n}}}$.\\
{\em Proof:} The fixed rate coding can be regarded as the special case of variable rate coding. Here, we can use the similar technique in the proof of Theorem 5. 

Let $C_n=(f_n,g_n)$ be a fixed rate code. For every codeword $b\in\mathcal{I}$, define $\mbox{BIN}_b=\{x^n\in\mathcal{X}^n: f_n(x^n)=b\}$. From the duality of Slepian-Wolf coding and channel coding, every $\mbox{BIN}_b$ can be seen as a set of channel code for the discrete memoryless channel defined by $P_{Y|X}$. So, we can upper bound $|\mbox{BIN}_b|$ like $|\mathcal{N}_{b,t,n}|$ in \eqref{eq59}. Following the similar derivation in previous section, we have
\begin{equation}
\label{eq75}
|\mbox{BIN}_b|\le e^{n(\ln|\mathcal{X}|-H(X|Y)-\delta_n+\frac{\ln2}{n}+\frac{-\ln\epsilon_n}{n})}
\end{equation}
where $\delta_n=\sigma_H(X|Y)\sqrt{\frac{\ln{n}}{n}}-\eta\sqrt{\frac{1}{n\ln{n}}}$.
Hence, 
\begin{eqnarray}
&&r(C_n) \nonumber \\
&\ge&\frac{1}{n}\ln\frac{|\mathcal{X}^n|}{\max_{b}|\mbox{BIN}_b|} \nonumber \\
&=&\frac{1}{n}\ln\frac{e^{n\ln|\mathcal{X}|}}{\max_b|\mbox{BIN}_b|} \nonumber \\
&\ge&H(X|Y)+\sigma_H(X|Y)\sqrt{\frac{\ln{n}}{n}} \nonumber \\
\label{eq76}
&&-\frac{\ln2}{n}-\frac{-\ln\epsilon_n}{n}-\eta\frac{1}{\sqrt{n\ln{n}}}
\end{eqnarray}
Substituting the expression of $\epsilon_n$ into \eqref{eq76}, we have
\begin{eqnarray}
r(C_n)&\ge&H(X|Y)+\sigma_H(X|Y)\sqrt{\frac{-2\ln\epsilon_n-\ln\ln{n}}{n}} \nonumber \\
\label{eq77}
&&-\frac{\ln2}{n}-\frac{-\ln\epsilon_n}{n}-\eta\frac{1}{\sqrt{n\ln{n}}}
\end{eqnarray}
Finally, based on \eqref{eq36}, we can conclude
\begin{eqnarray}
r(C_n)&\ge&H(X|Y)+\sigma_H(X|Y)\sqrt{\frac{-\ln\epsilon_n}{n}} \nonumber \\
&&-O\bigg(\frac{1}{n}\bigg)-O\bigg(\frac{-\ln\epsilon_n}{n}\bigg)-O\bigg(\frac{1}{\sqrt{n\ln{n}}}\bigg) \nonumber
\end{eqnarray} 
which proves the claim.


\begin{thebibliography}{1}

\bibitem{SW}D. Slepian and J. K. Wolf, "Noiseless coding of correlated information sources,Ó {\em IEEE Trans. on Inf. Th.}, vol. 19, pp. 471-80, 1973.

\bibitem{xiong}V. Stankovic, A.D. Liveris, Z. Xiong, C.N. Georghiades, "On code design for the Slepian-Wolf problem and lossless multiterminal networks," {\em IEEE Trans. on Inf. Th.}, vol. 52, pp. 1495-1507, 1973.

\bibitem{ramch}S.S. Prahan, K. Ramchandran, "Distributed source coding using syndromes (DISCUS): design and construction," in {\em Proc. of Data Compression Conference}, Mar 1999, Snowbird, UT, pp. 158-167.

\bibitem{rose}P. Koulgi, E. Tuncel, S. L. Regunathan, and K. Rose, ÒOn zero-error
coding of correlated sources,Ó {\em IEEE Trans. Inf. Theory}, vol. 49, no. 11, pp. 2856-2873, Nov. 2003.

\bibitem{he}D. He, L.A. Lastras-Montano, E. Yang; A. Jagmohan, J. Chen, "On the Redundancy of Slepian Wolf Coding," {\em IEEE Trans. Inf. Theory}, vol. 55, no. 12, pp. 5607-27, Dec. 2009.

\bibitem{yang}E. Yang, J. Meng, "Jar Decoding: Non-Asymptotic Converse Coding Theorems, Taylor-Type Expansion, and Optimality," available at http://arxiv.org/abs/1204.3658.

\bibitem{nep}E. Yang, J. Meng, "Non-asymptotic Equipartition Properties for Independent and Identically Distributed Sources," available at http://arxiv.org/pdf/1204.3661.pdf.

\bibitem{japan1}S. Kuzuoka, "On the redundancy of variable-rate Slepian-Wolf coding," in {\em Proc. of IEEE 2012 International Symposium on Information Theory and its Application (ISITA2012)}, Honolulu, HI, Oct. 2012, pp. 155-159.

\bibitem{japan2}S. Kuzuoka, ÒA simple technique for bounding the redundancy of source coding with side information,Ó in {\em Proc. of 2012 IEEE International Symposium on Information Theory (ISIT2012)}, Cambridge, MA, U.S.A., Jul. 2012, pp. 915-919.

\bibitem{type}I. Csiszar, "The Method of Types," {\em IEEE Transactions on Information Theory}, vol. 44, no. 6, pp. 2505-2523, Oct. 1998.

\end{thebibliography}
\end{document}